\documentclass[prd,a4paper,showpacs,showkeys,preprint,byrevtex]
{revtex4}
\usepackage{graphicx}
\usepackage{dcolumn}
\usepackage{amsmath}
\usepackage{bm}
\usepackage{epsfig}

\begin{document}

\title{Semirelativistic potential model for three-gluon
glueballs}

\author{Vincent \surname{Mathieu}}
\thanks{IISN Scientific Research Worker}
\email[E-mail: ]{vincent.mathieu@umh.ac.be}
\author{Claude \surname{Semay}}
\thanks{FNRS Research Associate}
\email[E-mail: ]{claude.semay@umh.ac.be}
\affiliation{Groupe de Physique Nucl\'{e}aire Th\'{e}orique,
Universit\'{e} de Mons-Hainaut,
Acad\'{e}mie universitaire Wallonie-Bruxelles,
Place du Parc 20, BE-7000 Mons, Belgium}
\author{Bernard \surname{Silvestre-Brac}}
\email[E-mail: ]{silvestre@lpsc.in2p3.fr}
\affiliation{LPSC Universit\'{e} Joseph Fourier, Grenoble 1 \\
CNRS/IN2P3, Institut Polytechnique de Grenoble, \\
Avenue des Martyrs 53, FR-38026 Grenoble-Cedex, France.}

\date{\today}

\begin{abstract}
The three-gluon glueball states are studied with the generalization of a semirelativistic
potential model giving good results for two-gluon glueballs. The Hamiltonian depends only on 3
parameters fixed on two-gluon glueball spectra: the strong coupling constant, the string
tension, and a gluon size which removes singularities in the potential. The Casimir scaling
determines the structure of the confinement. Our results are in good agreement with other
approaches and lattice calculation for the odderon trajectory but differ strongly from lattice
in the $J^{+-}$ sector. We propose a possible explanation for this problem.
\end{abstract}


\maketitle

\section{Introduction}
\label{sec:intro} 

The Quantum Chromodynamics (QCD) theory allows the existence of bound states
of gluons, called glueballs, but no firm experimental discovery of such states has been
obtained yet. Glueballs are important in the understanding of several mechanisms. Some authors
proposed that glueballs characterize de-confinement \cite{Shuryak:2001cp} and that their coupling
to proton affects the gluonic contribution to proton spin \cite{Kochelev:2005vd}.

An important difficulty is that glueball states might possibly mix strongly with nearby meson
states \cite{Kochelev:2005tu}. Nevertheless, the computation of pure gluon glueballs remains
an interesting task. This could guide experimental searches and provide some calibration for
more realistic models of glueballs.

Lattice calculations are undoubtedly a powerful tool to investigate the structure of
glueballs. A previous study \cite{morn99} predicts the existence of a lot of resonances
between 2 and 4~GeV. A recent update of this work \cite{chen06} confirms the results already
obtained. It is worth mentioning that recent lattice calculations confirm the hierarchy of the
glueball spectrum \cite{Meyer:2004gx}.

The potential model, which is so successful to describe bound states of quarks, is also a
possible approach to study glueballs
\cite{corn83,brau05,hou84,hou01,kaid06,Abreu:2005uw,Buisseret:2007de}. In a recent paper
\cite{brau04}, a semirelativistic Hamiltonian is used to compute two-gluon glueballs with
masses in good agreement with those obtained by lattice calculations of
Ref.~\cite{morn99}. This Hamiltonian, the model III in Ref.~\cite{brau04}, relies on the
auxiliary field formalism \cite{morg99,sema04} and on a one-gluon exchange (OGE) interaction
proposed in Ref.~\cite{corn83}. It depends only on three parameters: the strong coupling
constant $\alpha_S$, the string tension $a$, and a gluon size $\gamma$ which removes
singularities in the short-range part of the potential. The constituent gluon mass is
dynamically generated and it is assumed that the Casimir scaling determines the color
structure of the confinement. These two ingredients are actually necessary to obtain a good
agreement between the results from a potential model and from lattice calculations.

In a previous paper \cite{Mathieu:2006ggg}, we generalized the model built for two-gluon
systems in Ref.~\cite{brau04} for the low-lying ($L=0)$ spectrum three-gluon glueballs. The
purpose of this paper is to extend the results presented in Ref.~\cite{Mathieu:2006ggg} to higher
glueballs. Compared to previous models \cite{hou84,hou01}, our approach is characterized by
some improved features: semirelativistic kinematics, more realistic confinement, dynamical
definition of the gluon mass, consistent treatment of the gluon size. These points will be
detailed below. The masses of the lowest glueballs are computed with a great accuracy and
compared with lattice calculations \cite{morn99,chen06,Meyer:2004gx}. In Sec.~\ref{sec:ham},
the three-gluon Hamiltonian is built, and the structure of the studied glueballs is presented
in Sec.~\ref{sec:wf}. The three-gluon glueball spectrum is presented with the two-gluon
glueball spectrum from Ref.~\cite{brau04} and is discussed in Sec.~\ref{sec:res}. Some
concluding remarks are given in Sec.~\ref{sec:conc}.

\section{Hamiltonian}
\label{sec:ham}

\subsection{Parameters}
\label{ssec:param}

In Ref.~\cite{brau04}, two sets of parameters, denoted A and B, were
presented for the model III (see Table~\ref{tab:par}). With the set A,
it is possible to obtain glueball masses in agreement with the results
of some experimental works \cite{zou99,bugg00}: the lowest $2^{++}$
state near 2~GeV, the lowest $0^{++}$ state near 1.5~GeV, and the lowest
$0^{-+}$ state near 2.1~GeV. The values of $a$ and $\alpha_S$ are close
to the ones used in some recent baryon calculations \cite{naro02}. With
the set B, glueball masses were computed in agreement with the results
of the lattice calculations of Ref.~\cite{morn99}. If the
absolute glueball masses found in Ref.~\cite{brau04} with both sets are
strongly different, the relative spectra are nearly identical. As we use
in this work a three-body generalization of the Hamiltonian model III of
Ref.~\cite{brau04}, the two sets will also be considered. Nevertheless, 
in the following, we mainly 
focus our attention to the results obtained with the set B.

It is worth mentioning how the parameters have been determined in
Ref.~\cite{brau04}. The mass of the lightest $2^{++}$ is nearly
independent of the values of $\alpha_S$ and $\gamma$, but depends
strongly on $a$. So, this last parameter has been determined with this
$2^{++}$ state. The remaining parameters $\alpha_S$ and $\gamma$ have
then be
computed in order to reproduce the lightest $0^{++}$ and $0^{-+}$
states. The three states $2^{++}$, $0^{++}$, and $0^{-+}$ have been
chosen because they are possible experimental glueball candidates
\cite{zou99,bugg00} and because they are computed with relatively small
errors in lattice calculations \cite{morn99,chen06}.

\subsection{Confinement potential}
\label{ssec:conf}

A good approximation of the confining interaction between a quark and an
antiquark in a meson is given by the linear potential $a\, r$, where $r$
is the distance between the two particles and where $a$ is the string
tension. In a baryon, lattice calculations and some theoretical
considerations indicate that each quark generates a flux tube and that
these flux tubes meet in a junction point $\bm R_0$ which minimizes the
potential energy (the so-called Y-junction). Following this hypothesis,
the confinement in a baryon could be simulated by the three-body interaction
\begin{equation}
\label{vyqqq}
V_{qqq} = a \sum_{i=1}^{3} |\bm r_i- \bm R_0|,
\end{equation}
For such a potential, the point $\bm R_0$ minimizes also the
length of the three flux tubes and is identified with the Toricelli
point \cite{silv04}.

The energy density $\lambda_c$ of a flux tube (string tension) can
depend on the color
charge $c$ which generates it. Lattice calculations \cite{deld00} and
effective models of QCD \cite{sema04a} predict that the Casimir scaling
hypothesis is well verified in QCD, that is to say that the energy
density is proportional to the value of the quadratic Casimir operator
$\hat F_c^2$ of the colour source
\begin{equation}
\label{la}
\lambda_c = \hat F_c^2 \, \sigma.
\end{equation}
We have then $\lambda_q = \lambda_{\bar q} = 4\,\sigma/3 = a$ and
$\lambda_g = 3\,\sigma$.
In this work we will assume that the confinement
in a three-body color singlet is given by
\begin{equation}
\label{vyccc}
V_{ggg} = \sigma \sum_{i=1}^{3} \hat F_i^2\,|\bm r_i- \bm R_0|.
\end{equation}
This potential can be considered as the three-body generalization of the
confinement used in Ref.~\cite{brau04}. No constant potential is added,
contrary to usual Hamiltonians in mesons and baryons
\cite{corn83,simo01}.
Let us note that if the three color charges are not the same, $\bm R_0$
is no longer identified with the Toricelli point \cite{math06}.

Interaction~(\ref{vyccc}) is very difficult to use in a practical
calculation. A good approximation can be obtained for three identical
color charges by substituting $\bm R_0$ by the center of mass coordinate
$\bm R_{\text{cm}}$ and by renormalizing the potential by a factor $f$
which depends on the three-body system \cite{silv04}. For three
identical particles, the best value is $f=0.9515$. We will use this
approximation in the following, which seems more realistic than a
confinement obtained by the sum of two-body forces \cite{hou84,hou01}.

We include in our model the spin-orbit correction to the confinement
potential. It is given by the Thomas precession of the particles and
reads \cite{buiss07}
\begin{equation}\label{eq:LSconf}
   V^{LS}_{\text{Conf}} =  -\frac{1}{2\mu^2}\sum_{i=1}^3 \frac{1}{r_i}
   \frac{dV_{\text{Conf}}}{dr_i} \bm{L}_i\cdot\bm S_i.
\end{equation}
This second order correction depends on the effective gluon mass $\mu$. We review in
Sec.~\ref{ssec:dyn} the main feature of this effective mass. Let us note that a Y-junction for
the confinement leads to a three-body spin-orbit since the Torricelli point is function of the
three particles. But with our approximation relying on the center-of-mass, the Thomas precession 
term is a sum of three one-body spin-orbit interactions.

In Refs.~\cite{corn83,hou01,hou84}, the confinement potential
saturates
at large distances in order to simulate the breaking of the color flux
tube between gluons due to color screening effects. An interaction of
type~(\ref{vyccc}) seems a priori inappropriate since the potential
energy can grow without limit. But the phenomenon of flux tube breaking
must only contribute
to the masses of the highest glueball states. Moreover, it has been
shown that the introduction of a saturation could not be the best
procedure to simulate the breaking of a string joining two colored
objects \cite{swan05}.

\subsection{Dynamical constituent gluon mass}
\label{ssec:dyn}

Within the auxiliary field formalism (also called einbein field
formalism) \cite{morg99}, which can be considered as an approximate way
to handle semirelativistic Hamiltonians \cite{sema04,buis04}, the
effective QCD Hamiltonian has a kinetic part depending on the particle
\textbf{current} masses $m_i$ and the interaction is dominated by the
confinement. A \textbf{state-dependent constituent} mass
$\mu_i = \langle \sqrt{\bm p_i^2+m_i^2} \rangle$ can be defined for each
particle, and all relativistic corrections (spin, momentum, \ldots) to
the static potentials are then expanded in powers of $1/\mu_i$. This
approach has been used in Ref.~\cite{brau04} to build the two-gluon
Hamiltonian. So, the same formalism will be applied also in this paper.

Taking into account the considerations of Sec.~\ref{ssec:conf}, the
simplest generalization to a three-gluon system of the dominant part of
the model III two-gluon Hamiltonian of Ref.~\cite{brau04} is
\begin{equation}
\label{H0_3g}
H_0 = \sum_{i=1}^{3}\sqrt{\bm p^2_i} + \frac{3}{4} f \, a \sum_{i=1}^3
\hat F_i^2\,|\bm r_i- \bm R_{\text{cm}}|,
\end{equation}
with the condition $\sum_{i=1}^{3}\bm p_i = \bm 0$, since we work in the
center of mass of the glueball. The gluons have vanishing current
masses and their color is such that $\langle \hat F_i^2 \rangle=3$.
Contrary to some previous works \cite{corn83,hou01,hou84}, our
Hamiltonian is a
semirelativistic one. In Ref.~\cite{brau04}, it has been shown that it
is an important ingredient to obtain correct two-gluon glueball spectra.

Using the technique of Ref.~\cite{semay07}, it is
possible to obtain an analytical approximate formula giving the glueball
mass $M_0$ and the
constituent gluon mass $\mu_0$ (the three constituent gluon masses are
the same
since the wave function is completely symmetrized,
see Sec.~\ref{sec:wf}) 
\begin{equation}
\label{mu_3g}
M_0 \approx 6\, \mu_0 \quad \text{with} \quad
\mu_0 \approx \sqrt{f\,\sigma (N+3)}.
\end{equation}
$N=0, 1, \ldots$ is the excitation number.  With a value of the meson string tension
$a=4\, \sigma/3$ around 0.2~GeV$^2$, the smallest gluon constituent
mass is around 650~MeV. It is then relevant to use an expansion in
powers of $1/\mu_0$. Such a value of the gluon mass is in
agreement with the values used in Refs.~\cite{corn83,hou01,hou84},
but here the constituent mass is dynamically generated. 

Instead of using the auxiliary field formalism, it is possible to
consider relativistic corrections which are expanded in powers of
$1/E_i(\bm p_i)$ where $E_i(\bm p_i)=\sqrt{\bm p_i^2+m_i^2}$ (see for
instance Ref.~\cite{godf85}). But, this leads to very complicated
non local potentials which are difficult to handle.

\subsection{Short-range potential}
\label{ssec:sr}

The Hamiltonian $H_0$~(\ref{H0_3g}) gives the main features of the three-gluon glueball
spectra, but the introduction of a short-range potential is necessary to achieve a detailed
study. In Ref.~\cite{brau04}, a one-gluon exchange (OGE) interaction between two gluons,
coming from Ref.~\cite{corn83}, has been considered. It is not possible to use it directly for
a three-gluon glueball because the color structure of the interaction is different. So, we use
here the last version of a OGE interaction between two gluons developed specifically for
three-gluon glueballs \cite{hou84,hou01}. Its explicit form, which is very similar to the form
of the OGE interaction for two-gluon glueballs, is given below.

This interaction contains a tensor part and a spin-orbit part. In our previous study of the
low-lying spectrum \cite{Mathieu:2006ggg}, we neglected these terms since we worked on $L=0$
states. It has been shown that the tensor interaction between two gluons is small in two-gluon
glueballs \cite{brau04}. Hence, we only add a two-body spin-orbit interaction between each
pair of gluons.

The OGE two-gluon potential has a priori a very serious flaw: depending
on the spin state, the short-range singular part of the potential may be
attractive and lead to a Hamiltonian unbounded from below \cite{hou01}.
This problem is solved, as in Ref.~\cite{brau04}, by giving a finite
size to the gluon (see Sec.~\ref{ssec:size}).

The OGE potential depends on the gluon constituent mass. To
determine it, we follow the procedure proposed in Ref.~\cite{brau04}.
For a given set of quantum numbers $\{\alpha\}$, the eigenstate
$|\phi_\alpha\rangle$ of the Hamiltonian $H_0$ is computed. With this
state, a constituent gluon mass is computed
$\mu_\alpha = \langle \phi_\alpha |\sqrt{\bm p_1^2}|\phi_\alpha
\rangle$. This value of $\mu_\alpha$ is then used in the complete
Hamiltonian (see Sec.~\ref{ssec:ht}) to compute its eigenstate with
quantum numbers
$\{\alpha\}$. It is worth noting that, with this procedure, two states
which differ only by the radial quantum number are not orthogonal
since they are eigenstates of two different Hamiltonians which differ by
the value of $\mu$. It is shown in Ref.~\cite{sema04} that this problem
is not serious, the overlap of these states being generally weak.

\subsection{Gluon size}
\label{ssec:size}

In potential models, the gluon is considered as an effective degree of
freedom with a constituent mass. Within this framework, it is natural to
assume that a gluon is not a pure pointlike particle but an object
dressed by a gluon and quark-antiquark pair cloud. Such an
hypothesis for quarks leads to very good results in meson
\cite{brau98} and baryon \cite{brau02} sectors. As in
Ref.~\cite{brau04}, we assume here a Yukawa color charge density for the
gluon
\begin{equation}
\rho(\bm u)=\frac{1}{4\pi \gamma^2}
\frac{e^{-u/\gamma}}{u},
\label{dens}
\end{equation}
where $\gamma$ is the gluon size parameter.
The interactions between gluons are then modified by this density, a
bare potential being transformed into a dressed one.

The main purpose of the gluon dressing is to remove all singularities
in the short-range part of the interaction \cite{brau02}.
But, for consistency, the same regularization is applied to the
confinement potential, although no singularity is present in this case.
We think that the definition of a gluon size, which has a clear physical
meaning, is preferable to the use of
a smearing function only for potentials with singularity
\cite{brau05,hou01}.

A one-body potential, like $|\bm r_i-\bm R_{\text{cm}}|$, is dressed
by a simple convolution over the density of
the interacting gluon and the potential
\begin{equation}
\label{conv1}
V(\bm r)^*=\int d\bm r'\,V(\bm r')\, \rho(\bm r-\bm r').
\end{equation}
A dressed two-body potential, depending on $|\bm r_i-\bm r_j|$, is
obtained by a double convolution. This procedure is equivalent to the
following calculation
\cite{sema03}
\begin{equation}
\label{conv2}
V(\bm r)^{**}=\int d\bm r'\,V(\bm r')\, \Gamma(\bm r-\bm r')
\quad \text{with} \quad
\Gamma(\bm u)=\frac{1}{8\pi \gamma^3}e^{-u/\gamma}.
\end{equation}
Note that for the spatial dependencies of the spin-orbit
interactions, which depend on the derivative of the first order
potentials, the convolution must be performed before taking the
derivative.

For non vanishing value of $\gamma$, the value of the confinement potential at origin 
increases. This shifts the whole spectra to higher masses. Moreover, the strength of the 
Coulomb interaction is reduced. This also implies an increase of the glueball masses.

\subsection{Total Hamiltonian}
\label{ssec:ht}

To obtain the total Hamiltonian for three-gluon glueballs which is the simplest generalization
of the Hamiltonian for two-gluon glueballs from Ref.~\cite{brau04}, we take the Hamiltonian
$H_0$ given by the relation~(\ref{H0_3g}) and its spin-orbit correction \eqref{eq:LSconf} ; we
add the OGE interactions coming from Ref.~\cite{hou01} (without tensor part); and we dress all
the potentials with the gluon color density~(\ref{dens}). This gives the following Hamiltonian
($\hat F_i^2=3$, $\hat F_i\cdot\hat F_j=-3/2$)
\begin{subequations}
\label{h_3g}
\begin{eqnarray}
H &=& \sum_{i=1}^{3}\sqrt{\bm p^2_i}+ V_{\text{OGE}}^{**}+
V_{\text{Conf}}^* + V_{\text{Conf}}^{LS*}+ V_{\text{OGE}}^{LS**}\quad \text{with} \label{h_3ga}\\
V_{\text{OGE}}^{**} &=&-\frac{3}{2}\alpha_S \sum_{i<j=1}^3 
\left[\left(\frac{1}{4}+\frac{1}{3}\vec S\,^2_{ij}\right)
U(r_{ij})^{**} -\frac{\pi}{\mu^2}\delta(\bm r_{ij})^{**}
\left(\beta+\frac{5}{6}\vec S\,^2_{ij}\right)\right],\label{h_3gb}\\
U(r)^{**}&=&\frac{1}{(\mu^2\gamma^2-1)^2}\left(\frac{e^{-\mu r}}{r}
- \frac{e^{-r/\gamma}}{r}\right) +
\frac{e^{-r/\gamma}}{2\gamma(\mu^2\gamma^2-1)}
\quad \text{with} \quad U(r)=\frac{e^{-\mu r}}{r},\label{h_3gc}\\
\delta(\bm r)^{**}&=&\frac{1}{8\pi\gamma^3}e^{-r/\gamma},\label{h_3gd}
\\
V_{\text{Conf}}^*&=&\frac{9}{4} f \, a \sum_{i=1}^3 
y_i^* \quad \text{with} \quad y_i=|\bm r_i- \bm R_{\text{cm}}|
\quad \text{and} \quad 
r^*=r+2\gamma^2\frac{1-e^{-r/\gamma}}{r},\label{h_3ge}
\\
V_{\text{Conf}}^{LS*} &=& -\frac{9\,f\,a}{8\mu^2}\sum_{i=1}^3
  \bm{L}_i\cdot\bm S_i \frac{1}{y_i}
  \frac{d}{dy_i} y_i^* \\
V_{\text{OGE}}^{LS**} &=& -\frac{9\,\alpha_S}{4\mu^2}\sum_{i<j=1}^3
  \bm{L}_{ij}\cdot\bm S_{ij} \frac{1}{r_{ij}}\frac{d}{dr_{ij}}U(r_{ij})^{**}
\end{eqnarray}
\end{subequations}
where $\sum_{i=1}^{3}\bm p_i = \bm 0$ and $\vec S_{ij}=\vec S_i+\vec S_j$.
$\beta=+1$ ($-1$) for a gluon pair in color octet antisymmetrical
(symmetrical) state. The constituent state-dependent gluon mass $\mu$ is
computed in advance with a solution of the Hamiltonian $H_0$.

\section{Wave functions}
\label{sec:wf}

A gluon is a $I(J^P)=0(1^-)$ color octet state.
Two different three-gluon color singlet states exist \cite{hou84}, which
are completely symmetrical or completely antisymmetrical. The total
isospin state of a glueball is an isosinglet and is completely
symmetrical. Different total spin states are allowed with different
symmetry properties. They are presented in Table~\ref{tab:s}. As
gluons are bosons, the total wave function must be completely
symmetrical. Its parity is the opposite of the spatial parity, and its
$C$-parity is positive for
color antisymmetrical state and negative for color symmetrical state.
Let us note that a two-gluon glueball has always a positive $C$-parity.

In our previous work \cite{Mathieu:2006ggg}, we mainly considered glueballs with the lowest
masses. These states are characterized by a vanishing total orbital angular momentum $L=0$ and
by a spatial wave function completely symmetrical with a positive parity. This immediately
implies that the lowest glueballs are states with $J^{PC}$ equal to $0^{-+}$, $1^{--}$, and
$3^{--}$ \cite{hou84,hou01}. 
No firm conclusion could be drawn for the $0^{-+}$ state because it appears both 
in two-gluon and three-gluon spectra. Our results for the $1^{--}$ and
$3^{--}$ glueballs were in good agreement with lattice masses.  We also presented
the mass of the $2^{--}$ even though it was found with a mass higher than the one predicted by 
the lattice calculations. We could understand this bad
result as a first hint that possibly our approach could not be the best way to
handle pure gauge spectra.

We now extend our study to states with negative $C$-parity. Indeed, at least 3 gluons are contained into those
glueballs. Lattice QCD recently computed masses of $J^{+-}$ glueballs: $0^{+-} (4780), 1^{+-} (2980), 2^{+-}
(4230)$, $3^{+-} (3600)$ in Ref.~\cite{chen06}, and 
$1^{+-} (2670)$, $3^{+-} (3270)$, $5^{+-} (4110)$ in Ref.~\cite{Meyer:2004gx}. The positive
parity requests an odd angular momentum. 

In order to explain the high energy behavior of the proton-antiproton scattering,
the existence of a trajectory carrying vacuum quantum numbers was postulated: The
pomeron \cite{Donnachie:2002en}.
The matching of the two-gluon spectrum and this trajectory is an important success of the theory
\cite{Meyer:2004gx, pomeron}. The odderon is the negative $C$-parity counterpart of the
pomeron \cite{Lukaszuk:1973nt}. One often argues that the odd spin glueballs lie on its
trajectory. The first two states ($1^{--}$ and $3^{--}$) on its trajectory were already
computed in Ref.~\cite{Mathieu:2006ggg}. In this paper, we also computed the $5^{--}$ mass to
check this hypothesis. Although lattice do not report any result, we will compare our mass to
a Coulomb gauge approach for the odderon \cite{LlanesEstrada:2005jf}. The oddballs $J^{--}$
masses for $J=1,2,3,5,7$ as well as the $0^{-+}$ are computed within the Coulomb gauge
approach in this reference.

In order to reach a good accuracy, the trial spatial wave functions are expanded in large
Gaussian function basis \cite{suzu98}. Recently, we found the Fourier transform of a Gaussian
function and applied it to give the matrix elements for the semi-relativistic kinetic energy
\cite{SilvestreBrac:2007sg}. The interested readers may also find the matrix elements for spin
dependent operators in this base in Ref.~\cite{LS_tenseur}, where applications for three-body systems
were presented. With more than 10 Gaussian functions for each color/isospin/spin channels, we
have checked that the numerical errors on masses presented are around or less than 1~MeV.

Using the $a$ value from our previous models A and B \cite{brau04,Mathieu:2006ggg}, 
eigenvalues of the Hamiltonian $H_0$~(\ref{H0_3g}) are presented in
Table~\ref{tab:mu} for various $J^{PC}$ quantum numbers.
For each state, the corresponding constituent gluon mass $\mu_0$
is indicated. It is used to define the complete Hamiltonian
$H$~(\ref{h_3g}). Let us note that, even if the Hamiltonian $H_0$ does not contain
spin-dependent potential, the spin of the wave function can strongly influence the mass
through the symmetry of its spin-part. With our numerical procedure, 
the $J^{+\pm}$ glueballs are characterized by $L\ge 1$ spatial wave functions
with $L$ odd. In Table~\ref{tab:mu}, one can see that 
the four lowest $J^{+-}$ glueballs are degenerate ($L=1$, $S=1)$ as long as no
spin-dependent term is included in the potentials.

\section{Results}
\label{sec:res}

We present here the three-gluon glueball masses obtained with the complete Hamiltonian
$H$~(\ref{h_3g}) together with the two-gluon glueball masses computed in Ref.~\cite{brau04}
(see Table~\ref{tab:m}). These masses are compared with results obtained by the lattice
calculations of Ref.~\cite{chen06}. This last work is an update of a previous study \cite{morn99}.
So, states not computed in Ref.~\cite{chen06} but presented in Ref.~\cite{morn99} are also
considered here. The $5^{\pm -}$ masses are also computed and compared to the result 
of Ref.~\cite{Meyer:2004gx}.

This work is an extension of our previous study \cite{Mathieu:2006ggg} in which we concluded that our
$1^{--}$ and $3^{--}$ states were in good agreement with lattice results. However, our $2^{--}$ state is
clearly higher than its lattice counterpart. The lattice results predict a $2^{--}$ state at
4010~MeV near the $1^{--}$ and $3^{--}$ states. With our Hamiltonian, a mass more than 1~GeV
above is computed. It is unavoidable in our model, since a spin 2 function has a mixed
symmetry which implies a mixed symmetry for the space function and then a greater mass for the
corresponding glueball, in agreement with the results of Refs.~\cite{hou84,hou01}. It has been
checked that the spatial wave function of the $2^{--}$ state is dominated by a configuration
in which each internal variable is characterized by one unit of angular momentum. However in
Ref.~\cite{LlanesEstrada:2005jf}, the $2^{--}$ state lies between the $1^{--}$ and $3^{--}$. Actually,
they computed the spin-dependent corrections in perturbation with a wrong wave function. When
the correct wave function is used the $2^{--}$ goes higher in agreement with our model
\cite{private_comm}.

Moreover in our model, as long as we do not include spin-orbit forces the $0^{+-}$, $1^{+-}$,
$2^{+-}$, and $3^{+-}$ glueballs are degenerate $L=1$ states with the corresponding mixed
symmetry for the spin $S=1$ or 2 (see Table~\ref{tab:s}). We included spin-orbit interactions (coming
from the one-gluon exchange and Thomas precession) to split the degeneracy. But firstly, the
splitting between these states is not sufficient. And secondly, the hierarchy is not correct.
Indeed, we checked that the two spin-orbit contributions canceled roughly each other. The one
coming from the OGE decreases the masses by 100-200~MeV when the one coming from the
confinement increases the masses by around the same amount. This fact is well known in baryons
\cite{Isgur:1978xj} and hence it is not very surprising in a glueball potential model. 
Let us note that
this cancellation between two spin-orbit contributions was already noted for two-gluon glueballs 
\cite{brau04,kaid06}, but the effect coming from OGE was always stronger. 
In lattice calculations however, the situation is strongly different: The splitting between
$1^{+-}$ and $0^{+-}$ is about 1.8~GeV! Such a splitting cannot be reproduced in our simple
model. 

The values of parameters $a$ and $\alpha_S$ are quite well determined with lattice calculations 
and Regge phenomenology. The situation is very different for the gluon size $\gamma$. In this work,
it is assumed that $\gamma$ has the same value in two-gluon and three-gluon glueballs. We have checked that 
variations of the gluon size for three-gluon glueballs does not change the hierarchy of states, but the 
masses are globally shifted. As expected, the masses increase with $\gamma$.  

It is worth noting that the $1^{+-}$ and $3^{+-}$ three-gluon glueballs have masses similar to 
$J=2$ and $J=3$ two-gluon glueballs. Such low masses are difficult to explain. It is possible that 
these states are actually two-glueball bound states \cite{morn99}. Let us also note that $1^{+-}$ and $3^{+-}$ 
glueballs have low masses in the closed flux tube model \cite{Isgur85}.

We have no firm explanation for such discrepancies. These problems could arise because the
gluon has a constituent mass within our formalism. So, it possesses a spin as any massive
particle, that is to say three states of polarization, and we use a basis where each state is
labeled by a couple $(L,S)$. In the two-gluon sector, the $(L,S)$-basis leads to more states that
the ones predicted by the lattice and also to $J=1$ states forbidden by Yang's theorem. A
solution to cancel the extra states is to deal with the so-called helicity
formalism \cite{Jacob:1959at}. Recently, we reviewed and applied this formalism to two-gluon
glueballs \cite{Mathieu:2008bf}. We showed that a simple Cornell potential together with the
helicity formalism reproduces the correct hierarchy given by the lattice. An instanton induced
forced was also needed to raise the degeneracy between the scalar and pseudoscalar glueballs.  In
lattice calculations, the gluon is a massless particle with a definite helicity and then only
two states of polarization. We suspect that the implementation of the helicity formalism for
three-body systems would solve the hierarchy problem in the $J^{+-}$ sector but also for the
$2^{--}$. This difficult task is out of the scope of this paper and we leave it for future
work. Nevertheless, a attempt to generalization of the helicity formalism for three-body
system is presented in Ref.~\cite{Giebink:1985zz,stad97}.

It has been suggested that a three-body interaction can inverse the state ordering in the gluelump sector
\cite{Guo:2007sm}. It is not clear that such a force acts also in glueballs and could solve the hierarchy problem.
Actually, the dominant three-body force in glueball is the confinement, supplemented by its spin-orbit correction. 
In this work, as explained above, the confinement and its spin-orbit correction are 
simulated with a sum of one-body interactions. 

The odderon trajectory carries the quantum numbers $J^{--}$ with odd $J$. This trajectory was
investigated in Ref.~\cite{LlanesEstrada:2005jf} within a Coulomb gauge Hamiltonian formalism. 
In this work, states are built with well defined $(L,S)$ quantum numbers. 
The spectrum is in good agreement with our results (see Table~\ref{statetable}). 
Let us note that the Coulomb gauge Hamiltonian
is more complicated than ours. For instance, the annihilation diagram for two gluons is taken
into account whereas it vanishes in a potential model \cite{Mathieu:2007mw}. It is not
surprising that both approaches gives rather the same results since they used the same
formalism (a $(L,S)$-basis). We found that the two first oddballs $1^{--}$ and $3^{--}$ lie in
lattice error bars. This fact can be surprising at first glance. Indeed, we invoked the helicity
formalism as a possibility to solve the mass problem of the $J^{+-}$. Hence the same
problem should arise for the $J^{--}$. In the helicity formalism, a given $J^{PC}$ is a
particular combination of $(L,S)$ couples \cite{Jacob:1959at,Mathieu:2008bf}. We suspect than
the oddballs are largely dominated by the component $(L=J-3, S=3)$. It would be interesting to
have more information about those states but unfortunately lattice have difficulties to
identify higher excited states.

\section{Conclusion}
\label{sec:conc}

The masses of pure three-gluon glueballs have been studied with the
generalization of a semirelativistic potential model for pure two-gluon 
glueballs \cite{brau04}. The short-range part of the
potential is the sum of two-body OGE interactions. For the confinement,
a potential simulating a genuine Y-junction is used and it is assumed
that the Casimir scaling hypothesis is well verified. The gluon is
massless but the OGE interaction is expressed in terms of a state-dependent 
constituent mass. The Hamiltonian depends only on 3~parameters
fixed in Ref.~\cite{brau04}: the strong coupling constant $\alpha_S$, the string
tension $a$, and a gluon size $\gamma$. All masses have been accurately computed with
an expansion of trial states in Gaussian functions 
\cite{suzu98,SilvestreBrac:2007sg,LS_tenseur}.

>From our previous paper \cite{brau04,Mathieu:2006ggg}, we know that 
$J^{\pm +}$ two-gluon glueball spectra is in good agreement with
lattice calculations \cite{morn99,chen06}, but extra states not seen 
in lattice calculations are predicted. The $J^{-\pm}$ three-gluon glueball 
are also in quite good agreement with these lattice results, except for the 
$2^{--}$ state which is computed with a very high mass in our model.
In this work, the $5^{--}$ and some $J^{+-}$ three-gluon glueballs are finally computed. 
The $J^{--}$ candidates with $J$ odd for the odderon trajectory are in 
agreement with some other works. But, if the $0^{+-}$ and $2^{+-}$ 
states are predicted in quite good agreement with 
lattices results, it is not the case for the $1^{+-}$ and $3^{+-}$ states 
computed with too high masses in our model. One could interpret such a 
discrepancy as due to the fact that the spin-orbit forces are too feeble 
in our model to raise the degeneracy between $J^{+-}$ states. But we do not 
believe that a physical process, ignored here, is able to produce a strong enough 
spin-orbit force giving rise to more than 1~GeV energy gap. We think that the strong 
discrepancies between our results and lattice computations are due to the fact that
gluons do have constituent non vanishing masses
in our approach. They are then characterized by a spin, and not by a
helicity as it could be expected for particles with a vanishing current
mass. We have shown that working with the helicity formalism could cure the problem of
extra states in two-gluon glueballs \cite{Mathieu:2008bf}. We think that this formalism
applied to three-gluon glueballs could also improve the predictions of a potential model. 
Such a work is in progress.

\section*{acknowledgments}

The authors would like to thank Pedro Bicudo, Felipe Llanes-Estrada and F.~Buisseret for useful
discussions. V.~Mathieu (IISN Scientific Research Worker) and C.~Semay (F.R.S.-FNRS Research
Associate) would like to thank the F.R.S.-FNRS for financial support.


\clearpage

\begin{table}[h]
\protect\caption{Parameters for models A and B ($\sigma=3\,a/4$). For
both models, the gluon current mass is zero and $f=0.9515$.}
\label{tab:par}
\begin{ruledtabular}
\begin{tabular}{lll}
& Model A & Model B \\
\hline
$a$ & 0.16~GeV$^2$ & 0.21~GeV$^2$ \\
$\alpha_S$ & 0.40 & 0.50 \\
$\gamma$ & 0.504~GeV$^{-1}$ & 0.495~GeV$^{-1}$ \\
\end{tabular}
\end{ruledtabular}
\end{table}

\begin{table}[h]
\protect\caption{Characteristics of three-gluon spin functions with
total spin $S$, intermediate couplings $S_{\text{int}}$, and symmetry
properties which can be obtained by coupling
(A: Antisymmetrical, S: Symmetrical, MS: Mixed symmetry).
The multiplicity of each symmetry type is indicated. }
\label{tab:s}
\begin{ruledtabular}
\begin{tabular}{lll}
$S$ & $S_{\text{int}}$ & Symmetry \\
\hline
0 & 1 & 1 A \\
1 & 0, 1, 2 & 1 S, 2 MS \\
2 & 1, 2 & 2 MS \\
3 & 2 & 1 S
\end{tabular}
\end{ruledtabular}
\end{table}

\begin{table}[h]
\protect\caption{Masses $M_0$ of the Hamiltonian $H_0$~(\ref{H0_3g})
as a function of the $J^{PC}$ quantum numbers. The $(L,S)$ quantum
numbers are indicated with the corresponding
constituent gluon masses $\mu_0$. Values in MeV are
computed with the value of $a$ from models A/B. The lowest
masses are printed in italic.} \label{tab:mu}
\begin{ruledtabular}
\begin{tabular}{llllllll}
$J^{PC}$ & $(L,S)$ & $M_0$ & $\mu_0$ & $J^{PC}$ & $(L,S)$ & $M_0$ & $\mu_0$ \\
\hline
$0^{--}$ & (0,0) & 5574/6385 & 929/1064 & $0^{-+}$ & (0,0) & \emph{3211}/\emph{3679} &
535/613 \\
$1^{--}$ & (0,1) & \emph{3211}/\emph{3679} & 535/613 & $1^{-+}$ & (0,1) & 4156/4761 &
693/794
\\
$2^{--}$ & (0,2) & 4156/4761 & 693/794 & $2^{-+}$ & (0,2) & 4156/4761 & 693/794 \\
$3^{--}$ & (0,3) & \emph{3211}/\emph{3679} & 535/613 & $3^{-+}$ & (0,3) & 5574/6385 &
929/1064 \\
$5^{--}$ & (2,3) & 4182/4791 & 697/795\\
 \\
$0^{+-}$ & (1,1) & 3752/4298 & 625/717 \\
$1^{+-}$ & (1,1) & 3752/4298 & 625/717\\
$2^{+-}$ & (1,1) & 3752/4298 & 625/717\\
$3^{+-}$ & (1,2) & 3752/4298 & 625/717\\
$5^{+-}$ & (3,2) & 4596/5265 & 766/878

\end{tabular}
\end{ruledtabular}
\end{table}

\begin{table}[h]
\protect\caption{Glueball masses in MeV and (glueball mass ratios
normalized to lightest $2^{++}$). The two-gluon masses are taken from
Ref.~\cite{brau04}. The error bars for lattice mass ratios
are computed without the normalization error on the masses. The lightest
$0^{++}$, $2^{++}$, and $0^{-+}$ states are taken as inputs to fix the
parameters. The first column indicates the valence gluon content as
predicted by our model.}
\label{tab:m}
\begin{ruledtabular}
\begin{tabular}{lllll}
 &$J^{PC}$ & Lattice [Ref.] & Model A & Model B \\
\hline
$gg$&$0^{++}$ & 1710$\pm$50$\pm$80\phantom{00} ($0.72\pm0.03$)
\cite{chen06} & 1604 (0.78) & 1855 (0.78) \\
&& 2670$\pm$180$\pm$130 ($1.12\pm0.09$) \cite{morn99} & 2592 (1.26)
& 2992 (1.26) \\
&$2^{++}$ & 2390$\pm$30$\pm$120\phantom{0} ($1.00\pm0.03$)
\cite{chen06} & 2051 (1.00) & 2384 (1.00) \\
&$0^{-+}$ & 2560$\pm$35$\pm$120\phantom{0} ($1.07\pm0.03$)
 \cite{chen06} & 2172 (1.06) & 2492 (1.05)  \\
&         & 3640$\pm$60$\pm$180\phantom{0} ($1.52\pm0.04$)
 \cite{morn99} & 3228 (1.57) & 3714 (1.56)\\
&$2^{-+}$ & 3040$\pm$40$\pm$150\phantom{0} ($1.27\pm0.03$)
 \cite{chen06} & 2573 (1.25) & 2984 (1.25) \\
&         & 3890$\pm$40$\pm$190\phantom{0} ($1.63\pm0.04$)
 \cite{morn99} & 3345 (1.63) & 3862 (1.62) \\
&$3^{++}$ & 3670$\pm$50$\pm$180\phantom{0} ($1.54\pm0.04$)
 \cite{chen06} & 3132 (1.53) & 3611 (1.51) \\
 & & & & \\
$ggg$&$1^{--}$ & 3830$\pm$40$\pm$190\phantom{0} ($1.60\pm0.04$)
 \cite{chen06} & 3433 (1.67) & 3999 (1.68) \\
&$2^{--}$ & 4010$\pm$45$\pm$200\phantom{0} ($1.68\pm0.04$)
 \cite{chen06}& 4422 (2.16) & 5133 (2.15) \\
&$3^{--}$ & 4200$\pm$45$\pm$200\phantom{0} ($1.76\pm0.04$)
 \cite{chen06} & 3569 (1.74) & 4167 (1.75) \\
&$0^{-+}$ &  & 3688 (1.80) & 4325 (1.81)  \\
&$0^{+-}$ & 4780$\pm$60$\pm$230\phantom{0} ($2.00\pm0.05$)
 \cite{chen06} & 4043 (1.97) & 4656 (1.95)  \\
&$1^{+-}$ & 2980$\pm$30$\pm$140\phantom{0} ($1.25\pm0.03$)
 \cite{chen06} & 3992 (1.95) &  4626 (1.94) \\
&$2^{+-}$ & 4230$\pm$50$\pm$200\phantom{0} ($1.78\pm0.04$)
 \cite{chen06} & 3907 (1.90) &  4542 (1.91) \\
&$3^{+-}$ & 3600$\pm$40$\pm$170\phantom{0} ($1.51\pm0.04$)
 \cite{chen06} &  4033 (1.97) &  4568 (1.92)  \\
&$5^{+-}$ & 4110$\pm$170$\pm$190\phantom{0} \cite{Meyer:2004gx} & 4571 (2.23) & 5317 (2.23)  \\
&$5^{--}$ &  & 4521 (2.20) & 5263 (2.21)  \\
\end{tabular}
\end{ruledtabular}
\end{table}

\begin{table}[h]
\caption{\label{statetable} Odderon quantum numbers and  masses in MeV.
$L$ and $S$ assignments agree in our model B and in Ref.~\cite{LlanesEstrada:2005jf}.}
\begin{ruledtabular}
\begin{tabular}{ccccc}
$J^{PC}$& $1^{--}$ &  $3^{--}$ & $5^{--}$ & $7^{--}$
\\ \hline
$S$         & 1  &  3 & 3 & 3 \\
$L$         & 0  & 0 & 2 & 4 \\
\hline
Model B &  3999 & 4167 & 5263 &  \\
Coulomb Gauge \cite{LlanesEstrada:2005jf}&   3950 & 4150 & 5050 & 5900 \\
Lattice~\cite{chen06}&    3830  & 4200 & &  \\
Lattice~\cite{Meyer:2004gx}& 3100 & 4150 & &  \\
Wilson loops~\cite{ks}&  3490 &  4030 & & \\
\end{tabular}
\end{ruledtabular}
\end{table}

\clearpage

\begin{center}
\begin{figure}
\includegraphics*[height=8cm]{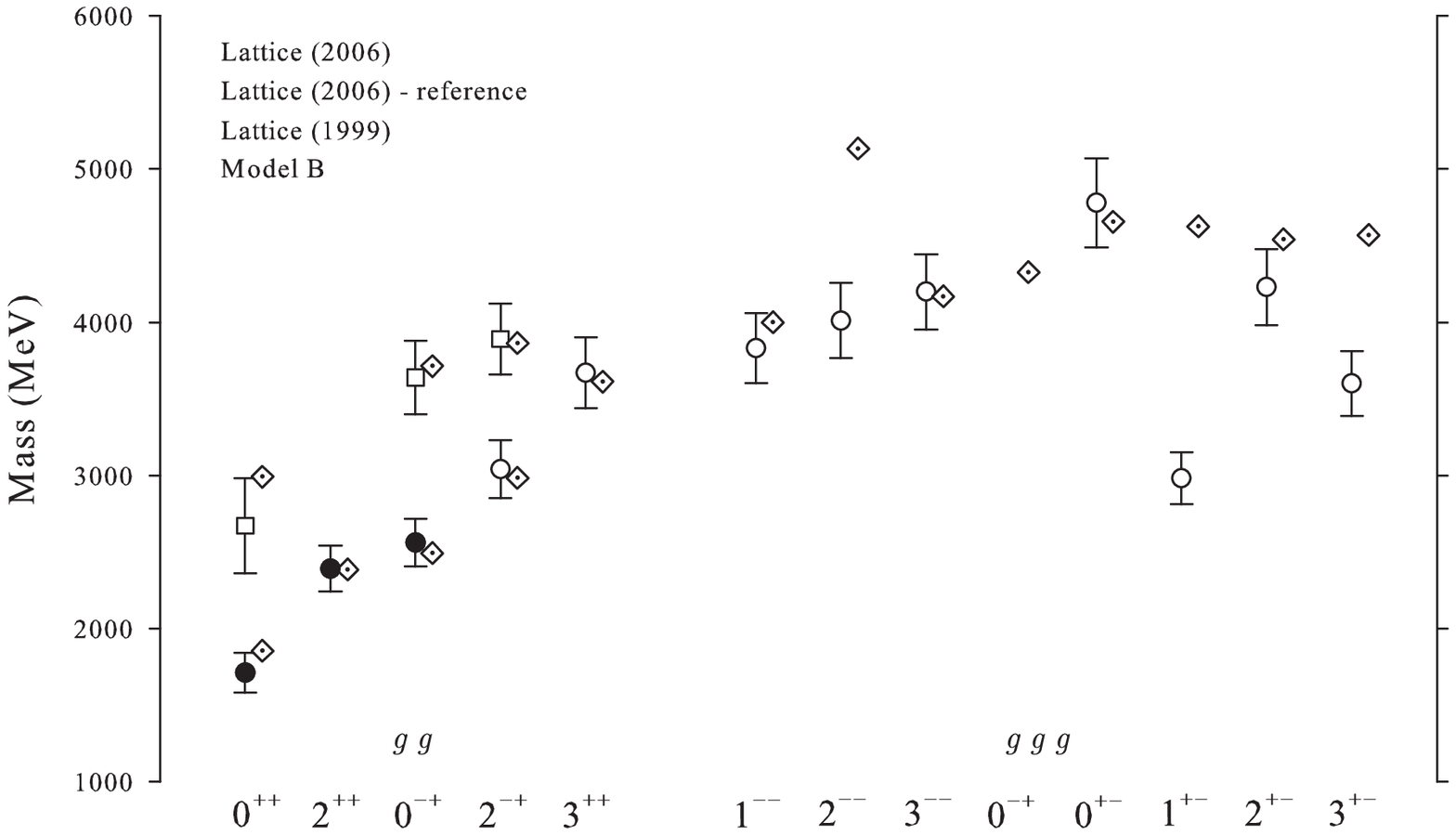}
\protect\caption{Glueball masses given in MeV. Dotted diamonds: Results
from model B (two-gluon masses are taken from
Ref.~\cite{brau04} and three-gluon $J^{-\pm}$ masses are taken from
Ref.~\cite{Mathieu:2006ggg}); Black and white circles: Lattice results from
Ref.~\cite{chen06}; White squares: Lattice results from
Ref.~\cite{morn99}. Black circles indicate the reference states taken as
inputs to fix the parameters. The error bars for lattice results are
computed by summing the two uncertainties (see Table~\ref{tab:m}).
}
\label{fig1}
\end{figure}
\end{center}


\begin{thebibliography}{aa}

\bibitem{Shuryak:2001cp}
  E.~V.~Shuryak,
  Phys.\ Lett.\  B {\bf 515} (2001) 359
  [arXiv:hep-ph/0101269].
  V.~Vento,
  Phys.\ Rev.\  D {\bf 75}, 055012 (2007)
  [arXiv:hep-ph/0609219].

\bibitem{Kochelev:2005vd}
  N.~Kochelev and D.~P.~Min,
  Phys.\ Lett.\  B {\bf 633}, 283 (2006)
  [arXiv:hep-ph/0508288].

\bibitem{Kochelev:2005tu}
  N.~Kochelev and D.~P.~Min,
  Phys.\ Rev.\  D {\bf 72}, 097502 (2005)
  [arXiv:hep-ph/0510016].
\bibitem{morn99} C. J. Morningstar and M. J. Peardon, Phys. Rev. D
{\bf 60}, 034509 (1999) [arXiv:hep-lat/9901004]. 
\bibitem{chen06} Y.~Chen \emph{et al.}, Phys. Rev. D
{\bf 73}, 014516 (2006) [arXiv:hep-lat/0510074]. 
\bibitem{Meyer:2004gx}
  H.~B.~Meyer,
  arXiv:hep-lat/0508002.
  H.~B.~Meyer and M.~J.~Teper,
   Phys.\ Lett.\  B {\bf 605}, 344 (2005)
  [arXiv:hep-ph/0409183].
\bibitem{corn83} J. M. Cornwall and A. Soni, Phys. Lett. B {\bf 120},
431 (1983).
\bibitem{brau05} F. Brau and C. Semay, Phys. Rev. D \textbf{72}, 078501
(2005) [arXiv:hep-ph/0411108]. 
\bibitem{hou84} W. S. Hou and A. Soni, Phys. Rev. D {\bf 29}, 101
(1984). 
\bibitem{hou01} W. S. Hou, C. S. Luo, and G. G. Wong, Phys. Rev. D
{\bf 64}, 014028 (2001) [arXiv:hep-ph/0101146]. 
\bibitem{kaid06} A. B. Kaidalov and Yu. A. Simonov,
Phys. Lett. B {\bf 636}, 101 (2006) [arXiv:hep-ph/0512151].

\bibitem{Abreu:2005uw}
  E.~Abreu and P.~Bicudo,
  J.\ Phys.\ G {\bf 34}, 195207 (2007)
  [arXiv:hep-ph/0508281].

\bibitem{Buisseret:2007de}
  F.~Buisseret,
  Phys.\ Rev.\  C {\bf 76}, 025206 (2007)
  [arXiv:0705.0916].

\bibitem{brau04} F. Brau and C. Semay, Phys. Rev. D \textbf{70}, 014017
(2004) [arXiv:hep-ph/0412173]. 
\bibitem{morg99} V. L. Morgunov, A. V. Nefediev, and Yu. A. Simonov,
Phys. Lett. B {\bf 459}, 653 (1999) [arXiv:hep-ph/9906318]. 

\bibitem{sema04} C. Semay, B. Silvestre-Brac, and I. M. Narodetskii,
Phys. Rev. D {\bf 69}, 014003 (2004) [arXiv:hep-ph/0309256]. 

\bibitem{Mathieu:2006ggg}
  V.~Mathieu, C.~Semay and B.~Silvestre-Brac,
  Phys.\ Rev.\  D {\bf 74}, 054002 (2006)
  [arXiv:hep-ph/0605205].

\bibitem{zou99} B. S. Zou, Nucl. Phys. {\bf A655}, 41 (1999). 
\bibitem{bugg00} D. V. Bugg, M. J. Peardon, and B. S. Zou, Phys. Lett. B
{\bf 486}, 49 (2000) [arXiv:hep-ph/0006179]. 
\bibitem{naro02} I. M. Narodetskii and M. A. Trusov,
Phys. Atom. Nucl. {\bf 65}, 917 (2002), Yad. Fiz. {\bf 65}, 949 (2002)
[arXiv:hep-ph/0104019];
Phys. Atom. Nucl. {\bf 67}, 762 (2004), Yad. Fiz. {\bf 67}, 783 (2004)
[arXiv:hep-ph/0307131].
\bibitem{silv04} B. Silvestre-Brac, C. Semay, I. M. Narodetskii, and A.
I. Veselov, Eur. Phys. J. C \textbf{32}, 385 (2004) [arXiv:hep-ph/0309247].
\bibitem{deld00} S.~Deldar, Phys. Rev. D {\bf 62}, 034509 (2000)
[arXiv:hep-lat/9911008]. G.~S.~Bali, Phys. Rev. D {\bf 62}, 114503 (2000)
[arXiv:hep-lat/0006022]. 
\bibitem{sema04a} C.~Semay, Eur. Phys. J. A {\bf 22}, 353 (2004)
[arXiv:hep-ph/0409105]. 
T.~H.~Hanson, Phys. Lett. B {\bf 166}, 343 (1986).
\bibitem{simo01} Yu. A. Simonov, Phys. Lett. {\bf 515}, 137 (2001)
[arXiv:hep-ph/0105141].
\bibitem{math06} V. Mathieu, C. Semay, and F. Brau, Eur. Phys. J. A
{\bf 27}, 225 (2006) [arXiv:hep-ph/0511210]. 
\bibitem{buiss07} F. Buisseret and C. Semay,
Phys. Rev. D {\bf 76}, 017501 (2007) [arXiv:0704.1745].
\bibitem{swan05} E.~S.~Swanson, J. Phys. G: Nucl. Part. Phys. {\bf 31},
845 (2005) [arXiv:hep-ph/0504097].
\bibitem{buis04} F. Buisseret and C. Semay,
Phys. Rev. D {\bf 70}, 077501 (2004) [arXiv:hep-ph/0406216]. 
\bibitem{semay07} C. Semay, F. Buisseret, N. Matagne, and Fl. Stancu,
Phys. Rev. D {\bf 75}, 096001 (2007) [arXiv:hep-ph/0702075].
\bibitem{godf85} S. Godfrey and N. Isgur, Phys. Rev. D {\bf 32}, 189
(1985).
\bibitem{brau98} F. Brau and C. Semay, Phys. Rev. D {\bf 58}, 034015
(1998) [arXiv:hep-ph/0412179]. 
\bibitem{brau02} F. Brau, C. Semay, and B. Silvestre-Brac,
Phys. Rev. C {\bf 66}, 055202 (2002) [arXiv:hep-ph/0412176]. 
\bibitem{sema03} B. Silvestre-Brac, F. Brau, and C. Semay,
J. Phys. G: Nucl. Part. Phys. {\bf 29}, 2685 (2003) [arXiv:hep-ph/0302252].

\bibitem{Donnachie:2002en}
  S.~Donnachie, H.~G.~Dosch, O.~Nachtmann and P.~Landshoff,
  Camb.\ Monogr.\ Part.\ Phys.\ Nucl.\ Phys.\ Cosmol.\  {\bf 19} (2002) 1.

\bibitem{pomeron}
  H.~B.~Meyer and M.~J.~Teper,
  Nucl.\ Phys.\ Proc.\ Suppl.\  {\bf 129}, 200 (2004)
  [arXiv:hep-lat/0308035].
  F.~J.~Llanes-Estrada, S.~R.~Cotanch, P.~J.~de A. Bicudo, J.~E.~F.~Ribeiro and A.~P.~Szczepaniak,
  Nucl.\ Phys.\  A {\bf 710}, 45 (2002)
  [arXiv:hep-ph/0008212].
  P.~Bicudo,
  arXiv:hep-ph/0405223.

\bibitem{Lukaszuk:1973nt}
  L.~Lukaszuk and B.~Nicolescu,
  Lett.\ Nuovo Cim.\  {\bf 8}, 405 (1973).

\bibitem{LlanesEstrada:2005jf}
  F.~J.~Llanes-Estrada, P.~Bicudo and S.~R.~Cotanch,
  Phys.\ Rev.\ Lett.\  {\bf 96}, 081601 (2006)
  [arXiv:hep-ph/0507205].

\bibitem{suzu98} Y. Suzuki and K. Varga, {\em Stochastic variational
approach to quantum mechanical few-body problems} (Springer Verlag,
Berlin, Heidelberg, 1998).
\bibitem{SilvestreBrac:2007sg}
  B.~Silvestre-Brac and V.~Mathieu,
  Phys.\ Rev.\  E {\bf 76}, 046702 (2007)
  [arXiv:0706.2300].
\bibitem{LS_tenseur}
  B.~Silvestre-Brac and V.~Mathieu, to be published in Phys. Rev. D, 
  arXiv:0712.0673.

\bibitem{bicu06} P. Bicudo, S. R. Cotanch, F. J. Llanes-Estrada, and D.
Robertson, arXiv:hep-ph/0602172.
\bibitem{bugg06} D. V. Bugg, arXiv:hep-ph/0603018.
\bibitem{abli06} M. Ablikim \emph{et. al} (BES Collaboration),
Phys. Rev. Lett. {\bf 96}, 162002 (2006) [arXiv:hep-ex/0602031].
\bibitem{private_comm} P. Bicudo and F.~J.~Llanes-Estrada, private communication.

\bibitem{Isgur:1978xj}
  N.~Isgur and G.~Karl,
  Phys.\ Rev.\  D {\bf 18}, 4187 (1978).

\bibitem{Isgur85}
  N.~Isgur and J.~Paton,
  Phys.\ Rev.\  D {\bf 31}, 2910 (1985).

\bibitem{Jacob:1959at}
  M.~Jacob and G.~C.~Wick,
  Annals Phys.\  {\bf 7}, 404 (1959)
  [Annals Phys.\  {\bf 281}, 774 (2000)].
\bibitem{Mathieu:2008bf}
  V.~Mathieu, F.~Buisseret and C.~Semay,
  arXiv:0802.0088.
\bibitem{Giebink:1985zz}
  D.~R.~Giebink,
  Phys.\ Rev.\  C {\bf 32}, 502 (1985).

\bibitem{stad97} A. Stadler, F. Gross, and M. Frank, 
Phys. Rev. C {\bf 56}, 2396 (1997). 

\bibitem{Guo:2007sm}
  P.~Guo, A.~P.~Szczepaniak, G.~Galata, A.~Vassallo and E.~Santopinto,
  arXiv:0707.3156.
  A.~P.~Szczepaniak and P.~Krupinski,
  Phys.\ Rev.\  D {\bf 73}, 034022 (2006)
  [arXiv:hep-ph/0511083].

\bibitem{Mathieu:2007mw}
  V.~Mathieu and F.~Buisseret,
  J.\ Phys.\ G {\bf 35}, 025006 (2008)
  [arXiv:hep-ph/0702226].

\bibitem{ks} A.~B.~Kaidalov and Yu.~A.~Simonov,
Phys. Lett. B \textbf{477}, 163 (2000) [arXiv:hep-ph/9912434].

\end{thebibliography}
\end{document}